\documentclass[aps,twocolumn,prl,showpacs,superscriptaddress,floatfix]{revtex4-1}
\usepackage{amssymb,amsmath,amsfonts}
\usepackage{epsfig,graphicx}

\begin{document}
\title{Non-thermal quantum phase transitions}

\author{Ricardo Puebla}
\affiliation{Grupo de F\'{\i}sica Nuclear, Departamento de F\'{\i}sica At\'omica, Molecular y Nuclear, Universidad Complutense de Madrid, Av. Complutense s/n, 28040 Madrid}

\author{Armando Rela\~no}
\affiliation{Departamento de F\'{\i}sica Aplicada I and GISC, Universidad Complutense de Madrid, Av. Complutense s/n, 28040 Madrid}

\begin{abstract}
  We report a kind of quantum phase transition which takes place in
  isolated quantum systems with non-thermal equilibrium states and an
  extra symmetry that commutes with the Hamiltonian for any values of
  the system parameters. A critical energy separates
  two different phases, one in which the symmetry is broken. This
  critical behavior is ruled out as soon as the system is put in
  contact with a thermal bath. The critical point is crossed when a
  sufficent amount of work is performed on the system, keeping it
  isolated from the environment. Different phases are identified by
  means of an order parameter, which is only different from zero in
  the symmetry-breaking phase. The behavior of the system near the
  critical point is determined by a set of critical exponents. We
  illustrate this phenomenon by means of numerical calculations in
  three different two-level systems.
\end{abstract}


\maketitle

{\em Introduction.-} Temperature is the fundamental magnitude in
equilibrium thermodynamics. Even for isolated systems, all the
thermodynamical information can be written in terms of the
microcanonical temperature $T=\left( k \partial S / \partial E
\right)^{-1}$. For example, a phase transition happens when the
behavior of the system abruptly changes at a certain critical
temperature, no matter whether the system is isolated or in contact
with a thermal bath. Notwithstanding, during the last couple of years
a number of isolated quantum systems with non-thermal equilibrium
states have been observed
\cite{Kinoshita:06,Hofferbeth:07,Gring:12}. It has been theoretically
proved that a final equilibrium state $\rho_{\text{eq}}$ is always
reached after a sufficently long-time evolution, meaning that for
almost any {\em reasonable} operator $A$, the time-dependent expected
value $\left\langle \Psi(t) \right| A \left| \Psi(t) \right\rangle$
remains close to $\text{Tr} \left[ \rho_{\text{eq}} A \right]$ for the
majority of times, independently of the initial condition $\left|
  \Psi(0) \right\rangle$ \cite{Reimann:12}. However, in this kind of
isolated quantum systems, the equilibrium state keeps large amounts of 
memory of
the initial condition, stored in a set of extra constants of motion,
and thus it has a non-thermal nature (see \cite{Polkovnikov:11} for a
recent review).

In this Letter we report a kind of quantum phase transition due to
this non-thermal behavior. In a class of quantum systems with a global and
discrete symmetry $S$, a transition from a symmetry-breaking to a
normal equilibrium state is observed at a certain critical energy
$E_c$, provided that this symmetry is broken in the initial
state. When this happens, and if the system stays isolated and recives
some energy in form of work, the set of extra constants of motion
determines whether the symmetry $S$ remains broken or is restored
after a sufficently long-time evolution. If some additional conditions
are fulfilled, this implies the existence of two different phases,
separated by a critical energy $E_c$ and characterized by an order
parameter, which is only different from zero in the symmetry-breaking
phase. Moreover, the behavior of the system around the critical energy
is determined by a set of critical exponents, which can be used to classify
these quantum phase transitions in different universality classes. 

{\em Nature of the quantum phase transition.-} The requisites for this
quantum phase transition are the following. First, a global and
discrete symmetry $S$ which commutes with the Hamiltonian for any
values of the system parameters, $\left[ H(\lambda),S \right]=0$,
$\forall \lambda$. Second, a standard quantum phase transition (QPT)
happening at $T=0$ for a certain critical value $\lambda_c$, that
distinguishes between two phases: one disordered and gapped, and the
other ordered and gapless, in which the symmetry $S$ can be
broken. Finally, an excited-state quantum phase transition (ESQPT) in
the ordered phase, which divides the spectrum in two different
regions: one with degenerated eigenvalues in which the symmetry $S$
can be broken, and another in which there are no degeneracies. This
kind of ESQPT have been recently studied in the Dicke model
\cite{Puebla:13}, and similar ones have been reported in a number of
models, covering quantum optics, molecular, atomic and nuclear physics
\cite{Cejnar:06,Caprio:08,Cejnar:08,Relano:08,Perez-Fernandez:11}.

\begin{figure}[h]
\centering
\includegraphics[width=0.45\textwidth,height=0.45\textwidth,angle=-90]{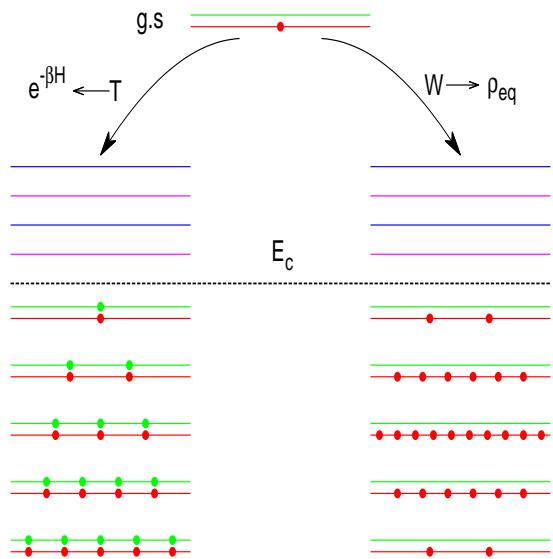}
\caption{Diagram illustrating the non-thermal nature of the quantum
  phase transition. Below $E_c$, red (green) lines represent
  symmetry-breaking energy levels with $\langle V \rangle = \eta$
  ($\langle V \rangle = - \eta$). Above $E_c$, blue (magenta) lines
  represent levels within positive (negative) subspaces of $S$.}
\label{Fig1}
\end{figure}

Let's consider an isolated quantum system that fulfills all the
previous requisites. Suppose that the symmetry $S$ is broken by the
action of a tiny external perturbation $\epsilon V$, so that the
double-degenerated ground state is splitted into a doublet, one level
characterized by $\left\langle V \right\rangle = \eta$, and the other
by $\left\langle V \right\rangle = - \eta$. The same structure is
propagated to the excited states up to a certain critical energy
$E_c$, above which the degeneracies are broken and all the eigenstates
have a well-defined value of $S$ (a diagram is plotted in
Fig. \ref{Fig1}). Now, freeze the system up to its ground state, the
lowest of the two levels of the doublet. Finally, give to the system
some amount of energy. If some heat is transferred by putting the
system in contact with a thermal bath at temperature $T$, the
conserved quantities of the Hamiltonian $H$ are ruled out, and its
eigenstates become populated according to the Boltzmann factor
$\exp(-\beta H)/Z$. This is sketched in left part of
Fig. \ref{Fig1}. On the contrary, if the energy is given in form of
work, keeping the system isolated from any environment, the
occupations of the eigenstates are determined by the extra conserved
quantities; a situation like the one depicted in the right part of
Fig. \ref{Fig1} occurs in the models considered below
\cite{footnote}. The differences between both cases are clear. In the
former, both levels of every doublet become occupied very
approximately with the same weight for any $T>\epsilon$; thus, the
average of every symmetry-breaking observable is zero, the symmetry
$S$ is restored and no phase transition is observed. On the other
hand, if the scenario depicted in right part of Fig. \ref{Fig1} holds,
the averages of $V$ and other observables which break the symmetry $S$
are different from zero. But this only happens below the critical
energy. For $E>E_c$, all the eigenstates have well-defined values of
$S$, so whatever their occupations in the final equilibrium
state are, the average of any symmetry-breaking observable is
zero. Therefore, the critical energy $E_c$ separates two different
phases, for which the nature of the equilibrium state $\rho_{
  \text{eq}}$ is qualitatively different.

{\em Physical models.-} We study this non-thermal quantum phase
transition in three different two-level models, for which the symmetry
$S$ is linked to the parity of the occupation of one of the levels. In
all the cases, the symmetry is changed just by promoting one particle
from the lower to the upper level (or vice versa), so it is easy to see
why a small thermal fluctuation breaks the symmetry and prevents the
ocurrence of this phase transition.

The two-mode Bose-Hubbard model (BH) describes a Bose-Einstein
condesate in a double-well potential \cite{Legget:01}
\begin{equation}
H=-J\left( a^{\dagger}_L a_R + a^{\dagger}_R a_L \right)
+ \frac{\lambda}{2N} \left[\hat{n}_L \left(\hat{n}_L-1 \right)+ \hat{n}_R \left(\hat{n}_R -1 \right) \right],
\end{equation}
where $\hat{n}_L$ $(\hat{n}_R)$ is the number operator,
$a^{\dagger}_L$ $(a^{\dagger}_R)$ and $a_L$ $(a_R)$ the usual creator
and annihilation operators for atoms at left (right) side of the well,
and $N=\hat{n}_L + \hat{n}_R$. The symmetry $S$ reflects the
invariance under the interchange between left and right
wells. Introducing $c_0^{\dagger}= \left( a_L^{\dagger} + a_R^{\dagger}
\right) / \sqrt{2}$ and $c_1^{\dagger} =\left( a_L^{\dagger} -
  a_R^{\dagger} \right) / \sqrt{2}$, $S$ is the parity of the number
of $c_0$ bosons, $S=\exp \left(i \pi c_0^{\dagger} c_0 \right)$. 

The Lipkin-Meshkov-Glick model (LMG) \cite{Vidal:06} describes the
interaction between two kinds of scalar bosons $s$ and $t$ bosons of
opposite parity,
\begin{equation}
H=\lambda t^{\dagger} t + \frac{1 - \lambda}{N} \left( s^{\dagger} t + t^{\dagger} s \right)^2,
\end{equation}
where $s^{\dagger}$ and $s$ are the usual creator and anihilation
operators for $s$ bosons; $t^{\dagger}$ and $t$, the same for $t$
bosons, and $N=s^{\dagger} s + t^{\dagger} t$ is the total number of
particles. $S$ is the parity of the number of $t$ bosons, $S = \exp
\left( i \pi t^{\dagger} t\right)$. 

The Dicke model (D) \cite{Dicke:54} describes the interaction between a
set of two-levels atoms and a single-mode radiation field,
\begin{equation}
  H = \omega_0 J_z + \omega a^{\dagger}a + \frac{2\lambda}{\sqrt{N}}\left(a^{\dagger}+a\right)J_x,
\label{eq:Dicke}
\end{equation}
where $a^{\dagger}$ and $a$ are the usual creation and annihilation
operators of photons, $\vec{J} = \left(J_x,J_y,J_z \right)$ is the
angular momentum, with a pseudo-spin length $J=N/2$, $N$ is the number
of atoms, $\omega$ is the frequency of the cavity mode, $\omega_0$ the
transition frequency, and $\lambda$ the coupling parameter.  In this
case, $S=\exp \left(i\pi \left[J+J_z+a^{\dagger}a \right] \right)$
\cite{Emary:03}. 

All these three models manifest both QPT and ESQPTs. The critical
energy $E_c$ can be estimated by means of a semiclassical
approximation, and lies on $E_c=-J$ for D \cite{Puebla:13,Perez-Fernandez:11},
$E_c=0$ for LMG \cite{Relano:08}, and $E_c=-\lambda/4 +1$ for BH
\cite{Julia:10}.
%  (see
% \cite{Julia:10} for BH, \cite{Relano:08} for L, and
% \cite{Hepp:73,Puebla:13} for D). 
It is worth to mention that both BH
and LMG have just one semiclassical degree of freedom, and thus they are
semiclassically integrable. However, D has two semiclassical
degrees of freedom and exhibits quantum and semiclassical chaos
\cite{Emary:03,Aguiar:92}. Therefore, albeit BH, LMG and D are two-level
models, they display different dynamics and can be classified into
different classes.

{\em Equilibrium state and order parameters.-} To characterize the
quantum phase transition, we need observables ${\mathcal O}$ for which
$\left\langle E_i \alpha \right| {\mathcal O} \left| E_i \alpha
\right\rangle=0$, where $\left| E_i \alpha \right\rangle$ denotes an
eigenstate with energy $E_i$ and a definite value of the symmetry $S$;
so if $\text{Tr} \left[ \rho_{eq} {\mathcal O} \right] \neq 0$, the
symmetry $S$ is broken in the equilibrium state. We choose
$Z=\hat{n}_L -\hat{n}_R=c_0^{\dagger} c_1 + c_1^{\dagger} c_0 $ for
BH, $s^{\dagger} t + t^{\dagger} s$ for LMG, and $J_x$ for D. In all the
cases, ${\mathcal O}$ changes a particle from one level into the
other. To study the behavior of the equilibrium states, we rely on the
following protocol:

{\it i)} Start from an initial state $\left| \Psi(0) \right\rangle$
which is the symmetry-broken ground state of the Hamiltonian $H
(\lambda_i)$ in the ordered phase, at a certain initial value of 
the coupling constant $\lambda_i$. We choose coherent states, which
give an accurate description of the ground state in the
thermodynamical limit
\begin{subequations}
\begin{align}
%\begin{split}
&\left |  \Psi(0) \right\rangle_{\text{BH}} = e^{\sqrt{N} \left( \gamma_0 a_L^{\dagger} + \gamma_1 a_R^{\dagger} \right)} \left| 0 \right\rangle, \;
\gamma_0^2 + \gamma_1 ^2 =1; \\
&\left|\Psi(0) \right\rangle_{\text{LMG}} = e^{\sqrt{\frac{N}{1+\beta^2}} \left( s^{\dagger} + \beta t^{\dagger} \right)} \left| 0 \right\rangle; \\
&\left|\Psi(0) \right\rangle_{\text{D}} = \frac{e^{\nu^2/2}}{\left(1+\mu^2\right)^{J}}  e^{\mu J_+ + \nu a^{\dagger}}\left| J,-J \right\rangle \otimes \left| 0 \right\rangle.
\end{align}
\label{coherent}
\end{subequations}

{\it ii)} Carry out a quench $\lambda_i \rightarrow
  \lambda_f$.  This entails to perform a work $\left\langle W \left( \lambda_i,
    \lambda_f \right) \right\rangle$ over the system.

{\it iii)} Let the system evolve under the final Hamiltonian
  $H (\lambda_f)$ until it reaches the final equilibrium state
  $\rho_{eq}$. As it is pointed in \cite{Reimann:12}, for any observable $A$
\begin{equation}
  \begin{split}
    \text{Tr} \left[ \rho_{eq} A \right] &= \overline{\left\langle
        \Psi(t) \right| A \left| \Psi(t) \right\rangle} \equiv \\ &\equiv \lim_{T \rightarrow \infty} \frac{1}{T} \int_0^T dt \, \left\langle
        \Psi(t) \right| A \left| \Psi(t) \right\rangle.
  \end{split}
\end{equation}
In the region in which the eigenstates are degenerated, 
\begin{equation}
  \begin{split}
    \left\langle \Psi(t) \right| A \left| \Psi(t) \right\rangle = \sum_{ij \alpha \beta} C^*_{i \alpha} C_{j \beta} e^{-i (E_i - E_j) t} \left\langle E_i \alpha \right| A \left| E_j \beta \right\rangle,
  \end{split}
\end{equation}
where the indexes $\alpha$ and $\beta$ run over positive and negative
values of the symmetry $S$, and $C_{i \alpha} = \left\langle E_i
  \alpha \right| \left. \Psi(0) \right\rangle$. The long-time average
is obtained by supressing all the oscillatory terms, and therefore
\begin{equation}
\rho_{\text{eq}} = \sum_{i \alpha} \left| C_{i \alpha} \right|^2 \left| E_i \alpha \right\rangle \left\langle E_i \alpha \right|
+ \sum_{i \alpha \beta} C_{i \alpha} C^*_{i \beta} \left| E_i \alpha \right\rangle \left\langle E_i \beta \right|.
\label{rho_eq}
\end{equation}
On the contrary, in the region without degeneracies $\rho_{\text{eq}}$
reduces to the usual diagonal ensemble \cite{Polkovnikov:11}. Note
that in numerical calculations a criterion is needed to determine when the
two levels of a doublet are degenerated and $\rho_{\text{eq}}$ must be
calculated as in Eq. (\ref{rho_eq}), instead of following the diagonal
approximation. To avoid ambiguities, in all our numerical results we
have considered that degeneracies exist until the semiclassical
critical energy $E_c$ is reached. Small quantitative differences can
be observed when using other criteria, but the qualitative behavior
remains the same.

{\it iv)} Finally, study the results for $\text{Tr} \left[
  \rho_{\text{eq}} {\mathcal O} \right]$ in terms of the energy of the
final state.

\begin{figure}
  \includegraphics[width=0.35 \textwidth,height=0.45
\textwidth,angle=-90]{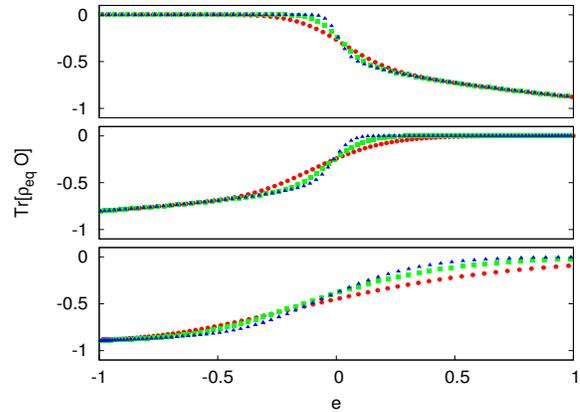}
\caption{Order parameter ${\mathcal O}$ for BH (top), LMG (middle) and
  D (bottom), in terms of the reduced energy
  $e=\varepsilon(E_f-E_c)/E_c$ and different system sizes, small (circles), medium (boxes) and large size (triangles) (see main text for details).}
  \label{order}
\end{figure}

In Fig. \ref{order} we plot $\text{Tr} \left[ \rho_{\text{eq}}
  {\mathcal O} \right]$ for BH with $\lambda_f = -7$ (upper panel),
LMG with $\lambda_f = 0.7$ (middle panel), and D with $\lambda_f =
0.75$ (lower panel), in terms of the reduced energy
$e=\varepsilon(E_f-E_c)/E_c$ and for three different system sizes.  In
all the cases $E_f$ and $E_c$ represent excitation energies. As the
spectrum of D is not bounded from above, we take $\varepsilon=1$ for
this case; thus $e$ is equivalent to the reduced temperature
$t=(T-T_c)/T_c$ of thermal phase transitions. On the contrary, the
specra of BH and LMG are bounded from above and the link to the
reduced temperature is not so clean. In consequence a scaling
parameter $\varepsilon \neq 1$ has been used just to make easier the
visual comparison between the three models. We have considered
$\hbar=1$, $J=1$ for BH, and $\omega_0=\omega=1$ for D, 
$N=500,2000$ and $8000$ particles for BH and LMG, and $N=16,32$ and
$64$ for D, since it requires much more computational resources.

The behavior of the expected value of the observables plotted in
Fig. \ref{order} clearly recalls the corresponding to the order
parameter of a phase transition. The expected value of ${\mathcal O}$
changes from non-zero in the ordered phase to zero in the normal one,
and the more particles we consider the more sudden change of
${\mathcal O}$ close to the critical energy is observed. We remark
that the behavior of D is smoother because it is obtained with a much
smaller number of particles, and that the ordered-phase happens for
$e<0$ in both LMG and D, whereas for $e>0$ in BH, due to the attractive
value of the interaction corresponding to the $\lambda_f$ chosen for
this model.

\begin{figure}
    \includegraphics[width=0.35 \textwidth,height=0.45
\textwidth,angle=-90]{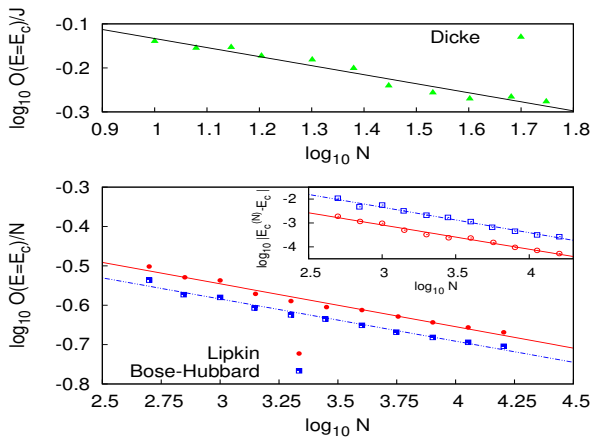}
  \caption{Expected value of the order parameter ${\mathcal O}$ at
  $E=E_c^{(N)}$ for D (top), BH (bottom, blue) and LMG (bottom, red)
  in function of the number of particles $N$. Inset of the bottom
  panel displays the scaling of $E_c^{(N)}$ for BH (blue) and LMG
  (red) in function of $N$.}
    \label{exp}
  \end{figure} 

  {\em Critical exponents and finite-size scaling.-} An exhaustive
  study of the order parameter near the critical energy is mandatory
  to determine if the behavior plotted in Fig. \ref{order} is the
  signature of a phase transition. Borrowing the language of
  second-order thermal phase transitions, we postulate that the
  trend of $\text{Tr} [\rho_{\text{eq}} {\mathcal O}]$ is universal
  around $E_c$, being each system characterized by a critical exponent
  $\beta$, $\text{Tr} [\rho_{\text{eq}} {\mathcal O}] \propto \left| E
    - E_c \right|^{\beta}$, for $E \sim E_c$. To obtain an estimate of
  this exponent for all the three models, we profit from the
  finite-size scaling of the critical energy. For a finite system of
  size $N$, the finite-size precursor of the critical energy
  $E_c^{(N)}$ scales as $\left| E_c^{(N)} - E_c \right| \propto
  N^{-\nu}$. So, at $E=E_c^{(N)}$ a finite-size scaling relation
  holds,
\begin{equation}
  \label{eq:exp_order} \text{Tr}
  [\rho_{\text{eq}} {\mathcal O}]_{E=E_c^{(N)}} \propto
  N^{-\zeta},
\end{equation} 
where $\zeta = \nu \beta$.  In Fig. \ref{exp} we represent the results
in double logarithmic scale as a function of the size system $N$. The
linear fit is also represented; its slope is directly the critical
exponent $\zeta$. The inset displays the behavior of $\lvert
E_c^{(N)}-E_c \rvert $ in order to obtain $\nu$ for LMG and BH (see
\cite{Puebla:13} for a calculation in the Dicke model). The resulting
critical exponents are summarized in Tab. \ref{Tab1}. We note that not
all the possible sources of error have been taken into account. For
example, changing the semiclassical $E_c$ by the finite-size precursor
$E_c^{(N)}$ in the calculation of $\rho_{\text{eq}}$ entails small
quantitative changes in $\text{Tr} [\rho_{\text{eq}} {\mathcal O}]$.
As it is needed an extra and very specific work to evaluate the
importance of all these factors, the results of Tab. \ref{Tab1} have
been obtained following exactly the same criterion. So, it is
reasonable to assume that all the uncontroled sources of error affect
equally to the three models, and thus a quantitative comparison
between them is possible. In any case, the errorbars should be
interpreted with caution.
\begin{table}
\begin{center}
  \begin{tabular}[c]{c| c | c|  c} \hline \hline Model  & $\zeta$   & $\nu$ & $\beta$ \\ \hline
BH & $\left(0.107 \pm 0.005\right)$ & $\left(1.07 \pm 0.04\right)$ & $\left(0.100 \pm 0.004\right)$ \\ 
LMG & $\left(0.109 \pm 0.005\right)$ & $\left(1.02\pm 0.04 \right)$ & $\left(0.107\pm 0.004\right)$ \\ 
D & $\left( 0.21\pm 0.01 \right)$ & $\left(1.30 \pm 0.02\right)$ & $\left(0.162 \pm 0.008\right)$ \\ \hline \hline
\end{tabular}
\end{center}
\caption{Critical exponents $\xi$, $\nu$, and $\beta$ for BH, LMG and D. Exponent $\nu$ for D is taken from ref. \cite{Puebla:13}.}
\label{Tab1}
\end{table}

These results entail two appealing outcomes. First, the neat power-law
scaling shown in Fig. \ref{exp} clearly suggest that the qualitative
change of $\rho_{\text{eq}}$ at the critical energy $E_c$ constitutes
a phase transition. In all the three models,
$\text{Tr}[\rho_{\text{eq}} {\mathcal O}]$ at $E=E_c^{(N)}$ goes to
zero as the thermodynamical limit is approached, following the same
kind of scaling of second-order thermal phase
transitions. Second, %  and despite extra work is required to obtain %
%precise estimates of the critical exponents,  
the results summarized in Tab. \ref{Tab1} allows us to conjecture 
that BH and LMG belong to
the same universality class, as both $\beta$ and $\nu$ exponents are
compatible. On the contrary, results for $D$ are significantly
different, and thus we can also conjecture that this model belongs to
a different universality class. This agrees with the fact that both BH
and LMG have just one semiclassical degree of freedom and are thus
integrable, whereas D has two semiclassical degrees of freedom and
manifests chaos. So, the same kind of non-thermal quantum phase
transition takes place in all these three models, but the precise
behavior of the order parameter near the critical energy depends on
the complexity of their dynamics.

{\em Conclusions.-} In this Letter we report a non-thermal quantum
phase transition due to the non-thermal nature of the equilibrium
states of a certain class of isolated quantum systems. This phase
transition takes place when some work is performed on a system with a
global symmetry $S$, provided that the system remains isolated from
the environment and the symmetry is broken in the initial state. It
entails an abrupt change in the equilibrium state at a
certain critical energy, as a consequence of the extra conserved
quantities of the Hamiltonian; so, it does not happen if the system is
put in contact with a thermal bath. We have shown this phenomenon in
three two-level quantum systems. We have found good order parameters
which characterize the different phases ---their averages in the
equilibrium states are zero in normal phase, and non-zero in the
ordered phase, where the symmetry is broken.  We have also defined a
set of critical exponents that characterize the behavior of the order
parameters close to the critical energy. We have obtained an estimate
of these exponents, from which we have conjectured that
Lipkin-Meshkov-Glick and Bose-Hubbard models belong to the same
universality class, while Dicke model belongs to a different one.

{\em Acknowledgments.-} This work is dedicated to the memory of
Joaqu\'{\i}n Retamosa. The authors acknowledge Marco Baity-Jesi for his
valuable comments. The work is supported in part by Spanish Government
grant for the research project FIS2009-07277. Part of the calculations
of this work were performed in the high capacity cluster for physics,
funded in part by Universidad Complutense de Madrid and in part with
Feder funding. This is a contribution to the Campus of International
Excellence of Moncloa, CEI Moncloa.


\begin{thebibliography}{99}

\bibitem{Kinoshita:06} T. Kinoshita, T. Wenger, and D. S. Weiss,
Nature {\bf 440}, 900 (2006).

\bibitem{Hofferbeth:07} S. Hofferbeth, I. Lesanovsky, B. Fischer,
T. Schumm, and J. Schmiedmayer, Nature {\bf 449}, 324 (2007).

\bibitem{Gring:12} M. Gring, M. Kuhnert, T. Langen, T. Kitagawa,
  B. Rauer, M. Screitl, I. Mazets, D. Adu Smith, E. Demler, and
  J. Schmiedmayer, Science {\bf 337}, 318 (2012).

\bibitem{Reimann:12} P. Reimann and M. Kastner, New. J. Phys {\bf 14}
(2012) 043020; A. J. Short, New. J. Phys {\bf 13} (2011) 053009.

\bibitem{Polkovnikov:11} A. Polkovnikov, K. Sengupta, A. Silva, and
M. Vengalattore, Rev. Mod. Phys. {\bf 83}, 863 (2011).

% \bibitem{Cazalilla:10} M. A. Cazalilla, and M. Rigol, New. J. Phys 
% {\bf 12} (2010) 055006.

\bibitem{Puebla:13} R. Puebla, A. Rela\~no, and J. Retamosa,
Phys. Rev. A {\bf 87}, 23819 (2013).

\bibitem{Cejnar:06} P. Cejnar, M. Macek, and S. Heinze, J. Phys. A
{\bf 39}, L515 (2006).

\bibitem{Caprio:08} M. A. Caprio, P. Cejnar, and F. Iachello,
Ann. Phys. (NY) {\bf 323}, 1106 (2008).

\bibitem{Cejnar:08} P. Cejnar and P. Str\'aansk\'y , Phys. Rev. E {\bf
78}, 031130 (2008).

\bibitem{Relano:08} A. Rela\~no, J. M. Arias, J. Dukelsky,
J. E. Garc\'{\i}a-Ramos, and P. P\'erez-Fern\'andez, Phys. Rev. A {\bf
78}, 060102(R) (2008); P. P\'erez-Fern\'andez, A. Rela\~no,
J. M. Arias, J. Dukelsky, and J. E. Garc\'{\i}a-Ramos, Phys. Rev. A
{\bf 80}, 032111 (2009).

\bibitem{Perez-Fernandez:11} P. P\'erez-Fern\'andez, P. Cejnar,
J. M. Arias, J. Dukelsky, J. E. Garc\'ia-Ramos, and A. Rela\~no,
Phys. Rev. A {\bf 83}, 033802 ( 2011); P. P\'erez-Fern\'andez,
A. Rela\~no, J. M. Arias, P. Cejnar, J. Dukelsky, and
J. E. Garc\'ia-Ramos, Phys. Rev. E {\bf 83}, 046208 (2011).

\bibitem{footnote}The symmetry-breaking
  term $\epsilon V$ rules out some of the conserved quantities of the
  system. However, as the contribution of this term is much smaller
  than the rest of the Hamiltonian, its only relevant effect is
  splitting the energy degeneracies. So, it makes sense to consider
  that all the original conserved quantities of the Hamiltonian remain
  unperturbed. All the results presented in this Letter are obtained
  without this symmetry-breaking term. Nevertheless, it has been
  checked that the results do not change if a small $\epsilon \sim
  10^{-6} - 10^{-8}$ is added.

\bibitem{Legget:01} A. J. Legget, Rev. Mod. Phys. {\bf 73}, 307
(2001).

\bibitem{Vidal:06} J. Vidal, J. M. Arias, J. Dukelsky, and
J. E. Garc\'{\i}a-Ramos, Phys. Rev. C {\bf 73}, 054305 (2006);
J. M. Arias, J. Dukelsky J. E. Garc\'{\i}a-Ramos, and J, Vidal, {\it
ibid.} {\bf 75}, 014301 (2007).

\bibitem{Dicke:54}   R. H. Dicke, Phys. Rev. {\bf 93}, 99 (1954).

\bibitem{Emary:03}   C. Emary, and T. Brandes, Phys. Rev. Lett. {\bf
90}, 044101 (2003); Phys. Rev. E {\bf 67}, 066203 (2003).

\bibitem{Julia:10} B. Juli\'a-D\'{\i}az, D. Dagnino, M. Lewenstein,
  J. Martorell, and A. Polls, Phys. Rev. A {\bf 81}, 023615 (2010).

\bibitem{Aguiar:92} M. A. M. de Aguiar, K. Furuya, C. H. Lewenkoff, and M. C. Nemes, Ann. Phys. (N. Y.) {\bf 216}, 291 (1992). 

\end{thebibliography}
\end{document}